% -------------------------------------------------------------------
% AIM  Assymptotic Iteration Method
% -------------------------------------------------------------------
% paper ciftci hall saad %
% it.tex -- it.tex [26 Aug 2003] [30 Sept 2003]
%
\magnification=\magstep1

% -------------------------------------------------------------------
%  generic unix 12 fonts (lower case names) with no magstep
% --------------------------------------------------------------------
\font\tr=cmr10                          % Our default
\font\bf=cmbx10                         % Redefinition
                         % Redefinition
\font\it=cmti10                         % Redefinition
\font\trbig=cmbx10 scaled 1500          % Main Title
          % large math italic bold 
                          % Theorems
\font\tiny=cmr8                         % Running title
% --------------------------------------------------------------------

% -------------------------------------------------------------------
\def\ptitle{Asymptotic iteration method for eigenvalue problems}
\output={\shipout\vbox{\makeheadline
                                      \ifnum\the\pageno>1 {\hrule}  \fi
                                      {\pagebody}
                                      \makefootline}
                   \advancepageno}
\headline{\noindent {\ifnum\the\pageno>1
                                   {\tiny \ptitle\hfil
page~\the\pageno}\fi}}
\footline{}
% ---------------------------------------------------------------------
\tr
%--------------------------------------------------------------------
    % bra ket:  math mode (to replace angle)
    %   ket  >
\def\nl{\hfil\break\noindent}  % new line after displayed equations
\def\ni{\noindent}             % noindent
\def\np{\hfil\vfil\break}
\def\ppl#1{{\noindent\leftskip 9 cm #1\vskip 0 pt}} % Preprint line 

 % bra < math mode
 % ket > math mode
\def\hi#1#2{$#1$\kern -2pt-#2} % hyphen \hi{N}{body} = N-body
\def\hy#1#2{#1-\kern -2pt$#2$} % hyphen hy{large}{N} = large-N

%--------------------------------------------------------------------

 % QED
 % SQUARE
%--------------------------------------------------------------------
% SPACING
% -------------------------------------------------------------------
\baselineskip 15 true pt  % draft 15
\parskip=0pt plus 5pt
\parindent 0.25in
\hsize 6.0 true in
\hoffset 0.25 true in
% 6 in width with 1.25 in margins default = (6.5, 0)
\emergencystretch=0.6 in                 % TEXBook p 107 : allows h-space
\vfuzz 0.4 in                            % page-length flexibility
\hfuzz  0.4 in                           % line-length flexibility
\vglue 0.1true in
\mathsurround=2pt                        % Default is 2pt
\topskip=24pt                            % Default is 10pt
% ---------------------------------------------------------------------
%  References
% ---------------------------------------------------------------------
\newcount\zz  \zz=0  % switch for printing references
\newcount\q   %  reference number
\newcount\qq    \qq=0  % starting reference number-1   (usually zero)
\def\pref#1#2#3#4#5{\frenchspacing \global \advance \q by 1     % paper reference
    \edef#1{[\the\q]}{\ifnum \zz=1{\item{$^{\the\q}$}{#2}{\bf #3},{ #4.}{~#5}\medskip} \fi}}
\def\bref #1#2#3#4#5{\frenchspacing \global \advance \q by 1     % book reference
    \edef#1{[\the\q]}
    {\ifnum \zz=1 { %
       \item{$^{\the\q}$}
       {#2}, {\it #3} {(#4).}{~#5}\medskip} \fi}}
\def\gref #1#2{\frenchspacing \global \advance \q by 1  % general reference
    \edef#1{[\the\q]}
    {\ifnum \zz=1 { %
       \item{$^{\the\q}$}
       {#2.}\medskip} \fi}}

 \def\sref #1{#1}
 
\def\references#1{\zz=#1
   \parskip=2pt plus 1pt   % default is 0pt plus 1pt
   {\ifnum \zz=1 {\noindent \bf References \medskip} \fi} \q=\qq
\pref{\nas}{N. Saad and R. L. Hall, J. Phys. A: Math. Gen. }{35}{4105 (2002)}{}

%-----------------------------------------------------------------------------
\bref{\grr}{G.E. Andrews, R. Askey and R. Roy}{Special Functions}{Cambridge: Cambridge University Press 1999}{}
%-----------------------------------------------------------------------------
\pref{\hald}{R. L. Hall, N. Saad and A. von Keviczky, J. Phys. A: Math. Gen. }{34}{1169 (2001)}{}
%-----------------------------------------------------------------------------
\pref{\hala}{R. L. Hall, N. Saad and A. von Keviczky, J. Math. Phys. }{39}{6345 (1998)}{}
%-----------------------------------------------------------------------------
\pref{\halb}{R. L. Hall and N. Saad, J. Phys. A: Math. Gen. }{33}{569 (2000)}{}
%-----------------------------------------------------------------------------
\pref{\halc}{R. L. Hall and N. Saad, J. Phys. A: Math. Gen. }{33}{5531 (2000)}{}
%--------------------------------------------------------------------
\pref{\harr}{E. M. Harrell, Ann. Phys. N.Y. }{105}{ 379 (1977)}{}
%-----------------------------------------------------------------------------
\pref{\agua}{V. C. Aguilera-Navarro, G.A. Est\'evez, and R. Guardiola, J. Math. Phys. }{31}{99 (1990)}{}
%-----------------------------------------------------------------------------
\pref{\aguc}{V. C. Aguilera-Navarro, F. M. Fern\'andez, R. Guardiola and J. Ros, J.Phys. A: Math. Gen. }{25}{6379 (1992)}{}

\pref{\agud}{V. C. Aguilera-Navarro, A. L. Coelho and Nazakat Ullah, Phys. Rev. A }{49} {1477 (1994)}{}
%-----------------------------------------------------------------------------
\pref{\zno3}{M. Znojil, Phys. lett. A}{158}{436 (1991)}{}
\pref{\zno4}{M. Znojil and P. G. L. Leach, J. Math. Phys. }{33}{2785 (1992)}{}
\pref{\sol}{W. Solano-Torres, G. A. Est\'eves, F. M. Fern\'andez,
and G. C. Groenenboom, J. Phys. A: Math. Gen. }{25}{3427 (1992)}{}
\pref{\half}{R. L. Hall, N. Saad and A. von Keviczky, J. Math. Phys. }{43}{94 (2002)}{}
\pref{\sim}{B. Simon, Ann. Phys.(N.Y.) }{58}{79 (1970)}{}
\pref{\bw}{C. M. Bender and T. T. Wu, Phys. Rev. }{184} {1231 (1969)}{}
\pref{\bms}{B. Bacus, Y. Meurice, and A. Soemadi, J. Phys. A: Math. Gen. }{28}{L381 (1995)}{}
\pref{\hall}{R. L. Hall, Canad. J. Phys. }{63}{311 (1985)}{}
 
\pref{\ali}{Ali Mostafazadeh, J. Math. Phys. }{42}{3372 (2001)}{}

}% end of ref list
\references{0}    % Initialization of reference numbers
% ------------------------------------------------------------------ end our ref.tex
% ----------------------------
% Preprint list
% ----------------------------
\ppl{CUQM-99}
\ppl{math-ph/0309066}
\ppl{September 2003}
%-------------------------------------------------------------------
%Title Page
%-------------------------------------------------------------------
%Title Page
%-------------------------------------------------------------------
\vskip 0.5 true in
\centerline{\bf\trbig\ptitle}
\medskip
\vskip 0.25 true in
\centerline{Hakan Ciftci$^*$, Richard L. Hall$^\dagger$ and Nasser Saad$^\ddagger$}
\bigskip
{\leftskip=0pt plus 1fil \rightskip=0pt plus 1fil\parfillskip=0pt
\obeylines $^*$Gazi Universitesi, Fen-Edebiyat Fak\"ultesi, Fizik
B\"ol\"um\"u, 06500 Teknikokullar, Ankara, Turkey.\par}
\medskip
{\leftskip=0pt plus 1fil
\rightskip=0pt plus 1fil\parfillskip=0pt
\obeylines
$^\dagger$Department of Mathematics and Statistics, Concordia University,
1455 de Maisonneuve Boulevard West, Montr\'eal,
Qu\'ebec, Canada H3G 1M8.\par}
\medskip
{\leftskip=0pt plus 1fil
\rightskip=0pt plus 1fil\parfillskip=0pt
\obeylines
$^\ddagger$Department of Mathematics and Statistics,
University of Prince Edward Island,
550 University Avenue, Charlottetown,
PEI, Canada C1A 4P3.\par}

\vskip 0.5 true in
%---------------------------------------------------------------------------
% Abstract
%---------------------------------------------------------------------------
\centerline{\bf Abstract}\medskip An asymptotic interation method for
solving second-order homogeneous linear differential equations of
the form $y^{\prime\prime}=\lambda_0(x)y^{\prime}+s_0(x)y$ is introduced, where
$\lambda_0(x)\neq0$ and $s_0(x)$ are $C_{\infty}$ functions. 
Applications to Schr\"odinger type
problems, including some with highly singular potentials, are presented.
\vfil\eject
%---------------------------------------------------------------------------
% 1. Introduction
%---------------------------------------------------------------------------
\ni{\bf 1. Introduction}
\medskip
\noindent Second-order homogeneous linear differential equations arise naturally in many fields in mathematical physics. There are many techniques available in the literature that can be used to solve these types of differential equation with boundary conditions. The main task of the present work is to introduce a new  technique, which we call the Asymptotic Iteration Method, to solve second-order homogeneous linear differential equations of the form
$$
y^{\prime\prime}=\lambda_0(x)y^\prime+s_0(x)y\eqno(1.1)
$$
where $\lambda_0(x)$ and $s_0(x)$ are defined in the some interval, not necessarily bounded, and $\lambda_0(x)$ and $s_0(x)$ have sufficiently many continuous derivatives. 
\medskip
%---------------------------------------------------------------------------
% 2. The Asymptotic Iteration Method
%---------------------------------------------------------------------------
\ni{\bf 2. The Asymptotic Iteration Method}
\medskip
\noindent Consider the homogenous linear second-order differential equation
$$y^{\prime\prime}=\lambda_0(x)y^\prime+s_0(x)y\eqno(2.1)$$
where  $\lambda_0(x)$ and $s_0(x)$ are functions in $C_{\infty}(a,b)$. In order to find a general solution to this equation we rely on the symmetric structure of the right hand side of (2.1). Indeed, if we differentiate (2.1) with respect to $x$, we find that
$$y^{\prime\prime\prime}=\lambda_1(x)y^\prime+s_1(x)y\eqno(2.2)$$
where
$$ \lambda_1= \lambda_0^\prime+s_0+\lambda_0^2,\hbox{ and } s_1=s_0^\prime+s_0\lambda_0.$$
If we write the second derivative of equation (2.1), we get
$$y^{\prime\prime\prime\prime}=\lambda_2(x)y^\prime+s_2(x)y\eqno(2.3)$$
where
$$ \lambda_2= \lambda_1^\prime+s_1+\lambda_0\lambda_1,\hbox{ and } s_2=s_1^\prime+s_0\lambda_1.$$
Thus, for $(n+1)^{th}$ and $(n+2)^{th}$ derivative, $n=1,2,\dots$, we have
$$y^{(n+1)}=\lambda_{n-1}(x)y^\prime+s_{n-1}(x)y\eqno(2.4)$$
and
$$y^{(n+2)}=\lambda_{n}(x)y^\prime+s_{n}(x)y\eqno(2.5)$$
respectively, where
$$ \lambda_{n}= \lambda_{n-1}^\prime+s_{n-1}+\lambda_0\lambda_{n-1},\hbox{ and } s_{n}=s_{n-1}^\prime+s_0\lambda_{n-1}.\eqno(2.6)$$
From the ratio of the $(n+2)^{th}$ and $(n+1)^{th}$ derivatives, we have
$${d\over dx}\ln(y^{(n+1)})={y^{(n+2)}\over y^{(n+1)}}=
{\lambda_n(y^\prime+{s_n\over \lambda_n}y)\over
\lambda_{n-1}(y^\prime+{s_{n-1}\over \lambda_{n-1}}y)}\eqno(2.7)$$
We now introduce the `asymptotic' aspect of the method. If we have, for sufficiently large $n$,
$${s_{n}\over \lambda_{n}}={s_{n-1}\over \lambda_{n-1}} := \alpha,\eqno(2.8)$$
\nl then (2.7) reduces to
$${d\over dx}\ln(y^{(n+1)})=
{\lambda_n\over \lambda_{n-1}}\eqno(2.9)$$
which yields
$$y^{(n+1)}(x)=
C_1\exp\bigg(\int\limits^x{\lambda_{n}(t)\over
\lambda_{n-1}(t)}dt\bigg) = C_1\lambda_{n-1}\exp\left(\int\limits^x(\alpha+\lambda_0)dt\right)\eqno(2.10)$$
where $C_1$ is the
integration constant, and the right-hand equation follows from (2.6) and the definition of $\alpha.$   Substituting (2.10) in (2.4) we obtain the
first-order differential equation
$$y^{\prime} + \alpha y = C_1\exp\left(\int\limits^x(\alpha+\lambda_0)dt\right)\eqno(2.11)$$
which, in turn, yields the general solution to (1.1)
as
$$y(x)= \exp\left(-\int\limits^{x}\alpha dt\right)\left[C_2 + C_1\int\limits^{x}\exp\left(\int\limits^{t}(\lambda_0(\tau) + 2\alpha(\tau)) d\tau \right)dt\right]
\eqno(2.12)$$
Consequently, we have proved the following theorem
\medskip
\noindent{\bf Theorem:} 

\noindent {\it Given $\lambda_0$ and $s_0$ in $C_{\infty}(a,b),$ then the differential equation
$$y^{\prime\prime}=\lambda_0(x)y^\prime+s_0(x)y$$
has a general solution (2.12)
if for some $n>0$
$${s_{n}\over \lambda_{n}}={s_{n-1}\over \lambda_{n-1}} \equiv \alpha,\eqno{(2.13)}$$
\nl where
$$ \lambda_{k}= \lambda_{k-1}^\prime+s_{k-1}+\lambda_0\lambda_{k-1}\hbox{ and }\quad s_{k}=s_{k-1}^\prime+s_0\lambda_{k-1}\eqno(2.14)$$
for $k=1,2,\dots,n$.}
\medskip
%--------------------------------
\ni{\bf 3. Some illustrative examples}
%--------------------------------
\medskip
%--------------------------------
\ni{\bf 3.1 Differential equations with constant coefficients}
%--------------------------------
\medskip
\noindent  If $s_0(x)$ and $\lambda_0(x)$ are constant
functions, for example, $\lambda_0 = 4$ and $s_0=-3$, then, for the differential equation 
$y^{\prime\prime}=4y^\prime-3y$, the computation of $\lambda_n$ and $s_n$ by means of Eq.(2.14) implies  
$$\lambda_n
={1\over 2} (3^{n+2} - 1)\quad \hbox{ and }\quad s_n = -{3\over 2} (3^{n+1} - 1).$$
The condition (2.13) implies  $\lim\limits_{n\rightarrow \infty}{\ s_{n}\over
\lambda_{n}}= -1$ which yields from (2.12) the general solution
$y(x)= C_2e^{x}+C_1e^{3x},$ as expected by the application of elementary methods. Generally speaking, if we consider the differential equation (1.1)
with $\lambda_0(x)$ and $s_0(x)$ are constants, we have from (2.14) that
$$\lambda_{n}=s_{n-1}+\lambda_0\lambda_{n-1}\hbox{ and }\quad s_{n}=s_0\lambda_{n-1}.$$
Consequently, the ratio ${s_{n}\over \lambda_{n}}$ becomes
$${s_{n}\over \lambda_{n}}={s_0\lambda_{n-1}\over s_{n-1}+\lambda_0\lambda_{n-1}}={s_0\over {s_{n-1}\over \lambda_{n-1}}+\lambda_0}$$
which yields, by means of (2.13), that
$${s_{n}\over \lambda_{n}}={s_0\over {s_n\over \lambda_n}+\lambda_0}$$
therefore
$$\bigg({s_n\over \lambda_n}\bigg)^2+\lambda_0{s_n\over \lambda_n}-s_0=0.\eqno(3.1)$$
This is a quadratic equation that can be used to find the ratio ${s_n\over \lambda_n}$ in terms of $\lambda_0$ and $s_0$.
Therefore, the expected solutions for the differential equation (1.1) with constant coefficients follow directly by means of (2.12).
\medskip
%--------------------------------
\ni{\bf 3.2 Hermite's differential equation}
%--------------------------------
\medskip
\noindent Many differential equations which are important in applications, such as the equations of Hermite, Laguerre, and Bessel, can be solve using the method discussed in Section~2. As an illustration we discuss here the exact solution of Hermite's Equation by means of the iteration method; other differential equations can be solved similarly.
Hermite's differential equation takes the form
$$f^{\prime\prime}=2xf^\prime-2kf,\quad -\infty<x<\infty. \eqno(3.2)$$
Here we have $\lambda_0=2x$ and $s_0=-2k$. Using (2.14), we can easily show that
$$\delta =\lambda_{n+1} s_{n}-s_{n+1}\lambda_{n}=2^{n+2}\prod\limits_{i=0}^{n+1}(k-i),\quad\quad n=0,1,2,\dots.\eqno(3.3)$$
Therefore, for the condition $\delta=0$ to hold, we must have $k$ a non-negative integer, usually known as the order of the Hermite equation.
Consequently, for each $k$, the ratio ${s_n\over \lambda_n}$ yields
$$k=0,\quad {s_0\over \lambda_0}={s_1\over \lambda_1}=\dots=0\Rightarrow f_0(x)=1$$
$$k=1,\quad {s_1\over \lambda_1}={s_2\over \lambda_2}=\dots=-{1\over x}\Rightarrow f_1(x)=x$$
$$k=2,\quad {s_2\over \lambda_2}={s_3\over \lambda_3}=\dots=-{4x\over 2x^2-1}\Rightarrow f_2(x)=2x^2-1$$
$$k=3,\quad {s_3\over \lambda_3}={s_4\over \lambda_4}=\dots=
-{6x^2-3\over 2x^3-3x}\Rightarrow f_3(x)=2x^3-3x$$
$$k=4,\quad {s_4\over \lambda_4}={s_5\over \lambda_5}=\dots=
{24x - 16x^3\over 3 -12x^2 + 4x^4}\Rightarrow f_4(x)=3 -12x^2 + 4x^4$$
and so on. Clearly, the expressions for the exact solutions $f_k(x)$  generate the well-known Hermite polynomials. We can easily verify that the general form of $f_k(x)$, $k=0,1,2,\dots$, is given  in terms of the confluent hypergeometric functions \sref{\grr} by
$$f_{2k}(x)=(-1)^k 2^k \left({1\over 2}\right)_k\ {}_1F_1\left(-k;\ {1\over 2};\ x^2\right),\eqno(3.4)$$
and
$$f_{2k+1}(x)=(-1)^k 2^k \left({3\over 2}\right)_k\ x\ {}_1F_1\left(-k;\ {3\over 2};\ x^2\right),\eqno(3.5)$$
where the Pochhammer symbol $(a)_k$ is defined by $(a)_0 = 1$ and $(a)_k = a(a+1)(a+2)\dots(a+k-1)$ for $k = 1,2,3,\dots,$ and may be expressed in terms of the Gamma function by $(a)_k={\Gamma(a+k)/ \Gamma(a),}$ when $a$ is not a negative integer $-m$, and, in these exceptional cases, $(-m)_k = 0$ if $k > m$ and otherwise $(-m)_k = (-1)^k m!/(m-k)!.$
\np %%% NEW PAGE KLUDGE %%%%%%%%%%%%%%%
%--------------------------------
\ni{\bf 3.3 Harmonic oscillator potential in 1-dimension}
%--------------------------------
\medskip\noindent
Although, the iteration method discussed in section 2 can be applied to any
second-order homogeneous linear differential equations of the form (1.1) with
$\lambda_0\neq0$, we shall concentrate in the rest of the article on the
eigenvalue problems of Schr\"odinger type. We shall show that Eq.(2.12) with the conditions (2.13)
and (2.14) gives a complete solution for many important Schr\"odinger-type problems. 
Through a concrete example we explore the exact solutions of Schr\"odinger's equation for the harmonic oscillator potentials, namely
$$
\left(-{d^2\over dx^2}+x^2\right)\psi=E\psi\eqno(3.6)
$$
where $\psi\in L_2(-\infty,\infty)$. In the limit of large $x$, the asymptotic solutions of (3.6) can be taken as any power of $x$ times a decreasing Gaussian. With this in mind we write the `unnormalized' wavefunctions as
$$\psi(x)=e^{-{x^2\over 2}}f(x)\eqno(3.7)$$
where the functions $f(x)$ to be found by means of the iteration procedure. Substituting (3.7) into (3.6), one obtains
$$
{d^2f\over dx^2}=2x {df\over dx}+(1-E)f\eqno(3.8)
$$
which, by comparison with (3.2), yields the exact eigenvalues
$$E_n=2n+1,\quad\quad n = 0,1,2,\dots$$
and the functions $f_n(x)$, $n=0,1,2,\dots$, are the Hermite polynomials obtained above. Therefore, using (3.4) and (3.5), the unnormalized wavefunctions of the Schr\"odinger equation (3.6) are: $$\psi_n(x)=(-1)^n 2^n \left({1\over 2}\right)_n\ e^{-{x^2\over 2}} {}_1F_1\left(-n;\ {1\over 2};\ x^2\right)\eqno(3.9)$$
for $n=0,2,4,\dots$
and 
$$\psi_n(x)=(-1)^n 2^n \left({3\over 2}\right)_n\ x e^{-{x^2\over 2}}{}_1F_1\left(-n;\ {3\over 2};\ x^2\right).\eqno(3.10)$$
for $n=1,3,\dots$. The normalization constant of $\psi(x)$ can be computed by means of $||\psi||=1,$ as we shall shortly show.
\medskip
\np %%% NEW PAGE KLUDGE %%%%%%%%%%%%%%%
%--------------------------------
\ni{\bf 3.4 Gol'dman and Krivchenkov Potential}
%--------------------------------
\medskip
\noindent The Gol'dman and Krivchenkov Hamiltonian is the generalization of the  Harmonic-oscillator Hamiltonian in 3-dimensions; namely
$$
\left(-{d^2\over dr^2}+r^2+{\gamma(\gamma+1)\over r^2}\right)\psi=E\psi\eqno(3.11)
$$
where $\psi\in L_2(0,\infty)$ and satisfies the condition $\psi(0)=0$ known as the Dirichlet boundary condition. The generalization lies in the parameter $\gamma$ ranging over [$0,\infty)$ instead of  the angular momentum quantum number $l=0,1,2,\dots$. For large $r$, the exact solutions of (3.11) are asymptotically equivalent to the exact solutions of the harmonic-oscillator problem with eigenfunctions vanishing at the origin. Therefore, we may assume that the unnormalized wavefunction $\psi$ takes the form
$$\psi(r)=r^{\gamma+1}e^{-{r^2\over 2}}f(r)\eqno(3.12)$$
where, again, $f(r)$ is to be determined through the iteration procedure discussed in Section~2. Substituting (3.12) into (3.11), we obtain
$$
{d^2f\over dr^2}=2\left(r-{\gamma+1\over r}\right) {df\over dr}+(2\gamma+3-E)f.\eqno(3.13)
$$
where $\lambda_0(r)=2(r-{\gamma+1\over r})$ and $s_0(r)={2\gamma+3-E}$. By means of Eq.(2.14) we may compute $\lambda_n(r)$ and $s_n(r).$ That result, combined with the condition (2.13), yields $$E_0=3+2\gamma,\quad E_1=7+2\gamma,\quad E_2=11+2\gamma,\quad \dots$$ respectively, that means
$$E_n=4n+2\gamma+3,\hbox{ for }\quad n=0,1,2,\dots.\eqno(3.14)$$ 
Furthermore, with the use of $f(r)=\exp(-\int{s_n\over \lambda_n}dr)$, Eq.(2.13), after some straightforward computations, yields 
$$f_n(r)=\sum\limits_{k=0}^n (-1)^k2^{n-2k}{\Gamma(n+1)\Gamma(2\gamma+2n+2)\Gamma(\gamma+n-k+1)\over \Gamma(k+1)\Gamma(n-k+1)\Gamma(n+\gamma+1)\Gamma(2n+2\gamma-2k+2)}r^{2n-2k}\eqno(3.15)$$
In order to show that (3.15) together with (3.12) yields the exact wavefunctions for the Gol'dman and Krivchenkov Hamiltonian, we may proceed as follows. Using the Pochhammer's identity $(a)_{-k}={\Gamma(a-k)/ \Gamma(a)}$, we can write (3.15) as
$$f_n(r)=2^n r^{2n}\sum\limits_{k=0}^n (-1)^k {(\gamma+n+1)_{-k}\over k!(n+1)_{-k}(2n+2\gamma+2)_{-2k}}\bigg({1\over 2r}\bigg)^{2k}$$
Since
$(a)_{-k}={(-1)^k/ (1-a)_k}$ we have 
$$f_n(r)=2^n r^{2n}\sum\limits_{k=0}^n  {(-n)_k(-n-\gamma-{1\over 2})_{k}\over k!}\bigg({1\over r^2}\bigg)^{k}\eqno(3.16)
$$
in which we have used Gauss's duplication formula 
$$(a)_{2n}=2^{2n}\bigg({a\over 2}\bigg)_n\bigg({a+1\over 2}\bigg)_n.$$ 
The finite sum in (3.16) is the series representation of the hypergeometric function ${}_2F_0$. Therefore
$$\eqalign{f_n(r)&=2^n r^{2n}\ {}_2F_0\left(-n,-n-\gamma-{1\over 2};-;-{1\over r^2}\right)\cr
&=(-1)^n2^n~n!~L_n^{\gamma+{1\over 2}}(r^2)
\cr
&=(-1)^n2^n\left(\gamma+{3\over 2}\right)_n{}_1F_1\left(-n;\gamma+{3\over 2};r^2\right)}$$
Consequently, the unnormalized wavefunctions take the form
$$\psi(r)=(-1)^n2^n\left(\gamma+{3\over 2}\right)_nr^{\gamma+1}e^{-{r^2\over 2}}{}_1F_1\left(-n;\gamma+{3\over 2};r^2\right)$$
for $n=0,1,2,\dots$. The normalization constant for $\psi(r)$ can be found using $||\psi||=1$ which leads to the exact wavefunctions of the Gol'dman and Krivchenkov potential, namely
$$\psi(r)=(-1)^n\sqrt{2(\gamma+{3\over 2})_n\over n!~\Gamma(\gamma+{3\over 2})}r^{\gamma+1}e^{-{r^2\over 2}}{}_1F_1\left(-n;\ \gamma+{3\over 2};\ r^2\right).\eqno(3.17)$$
Some remarks are in order:
\item{1.} The exact odd and even solutions of the harmonic oscillator potential in 1-dimension can be recovered from (3.17) by setting $\gamma=0$ and $\gamma=-1$ respectively.
\item{2.} The exact solutions  of the harmonic oscillator potential in 3-dimensions can be recovered from (3.17) by setting $\gamma=l,$ where $l=0,1,2,\dots$ is the angular momentum quantum number.
\item{3.} The exact solutions of the harmonic oscillator potential in $N$-dimensions can be recovered from (3.17) by setting $\gamma=l+{1\over 2}(N-3)$ where $N\geq 2$.
\medskip
\np %%% NEW PAGE KLUDGE %%%%%%%%%%%%%%%
%--------------------------------
\ni{\bf 4. Singular Potentials}
%--------------------------------
\medskip
We discuss in this section the application of the iteration method discussed in section 2 to investigate two important classes of singular potentials. The first class characterized by the generalized spiked harmonic oscillator potentials which has a singularity at the origin, and the second class characterized by the quartic anharmonic oscillator potentials where the perturbative term diverges strongly at infinity. Different approaches are usually applied to deal with each of these classes.  The asymptotic iteration method, however, can be used to investigate the eigenvalues for both classes of potentials.
\medskip

%--------------------------------
\ni{\bf 4.1 Generalized Spiked Harmonic Oscillator Potentials}
%--------------------------------
\medskip
\noindent Since the interesting work of Harrell \sref{\harr} on the ground-state energy of the singular Hamiltonian
$$H=-{d^2\over dx^2}+x^2+{A\over x^\alpha},\quad x\in [0,\infty),~ A\geq 0,~ \alpha>0,\eqno(4.1)$$
known as the spiked harmonic oscillator Hamiltonian, the volume of research in this field has grown rapidly. A variety of techniques have been employed in the study of this interesting family of quantum Hamiltonians \sref{\hala-\half}. We shall investigate here the solutions of the spiked harmonic oscillator Hamiltonian (4.1) in arbitrary dimensions by means of the iteration method. That is to say, we examine the eigenvalues of the Hamiltonian known as the generalized spiked harmonic oscillator Hamiltonian
$$H=-{d^2\over dx^2}+x^2+{\gamma(\gamma+1)\over x^2}+{A\over x^\alpha},\quad x\in [0,\infty),~ A\geq 0,~ \alpha>0,\eqno(4.2)$$
where $\gamma=l+(N-3)/2$ for $N\geq 2$. Clearly we may compute the eigenvalues of the spiked harmonic oscillator Hamiltonian (4.1) directly by setting $N=3$ and $l=0$ in (4.2). 

The wavefunctions (3.12) of the Gol'dman and Krivchenkov Hamiltonian suggests that the exponent in the power of $x$ term of the exact solutions of (4.2) should depend, at least, on the parameters $\gamma$ and $A$. Since the exact form of this term is unknown, we write the exact wavefunctions in the simpler form 
$$\psi(x)=e^{-{x^2\over 2}}f(x),\eqno(4.3)$$
where the functions $f(x)$ must satisfy the condition $f(0)=0$ and remain to be determined by the iteration method. Substituting (4.3) into (4.2), we obtain
$$
{d^2f\over dx^2}=2x {df\over dx}+\left(1-E+{A\over x^\alpha}+{\gamma(\gamma+1)\over x^2}\right)f.\eqno(4.4)
$$
With $\lambda_0(x)=2x$ and $s_0(x)=1-E+{A\over x^\alpha}+{\gamma(\gamma+1)\over x^2}$, we may compute  $\lambda_n(x)$ and $s_n(x)$ using (2.14) and the eigenvalues may be calculated by means of the condition (2.13). For each iteration, the expression $\delta= s_{n}\lambda_{n-1}-s_{n-1}\lambda_{n}$ will depend on two variables $E$ and $x$. The eigenvalues $E$ computed by means of $\delta=0$ should, however, be independent of the choice of $x$. Actually, this will be the case for most iteration sequences. The choice of $x$ can be critical to the speed of the convergence to the eigenvalues, as well as for the stability of the process. Although, we don't have at the moment a specific method to determine the best initial value of $x$, we may suggest the following approaches. For the spiked harmonic oscillator potential $V(x)=x^2+{A\over x^\alpha}$, one choice of $x=x_0$ could be the value of $x_0$ that minimize  the potential $V$, namely $x_0=({\alpha A\over 2})^{1/(\alpha + 2)}$. Another possible choice comes from noticing that the ground-state energy of the harmonic oscillator or that of the Gol'dman and Krivchenkov Potential can be obtained by setting $s_0=0$. We may therefore start our iteration with $x_0$ that is obtained from $s_0=0$. For example, if $\alpha=4$, then
$$s_0= 1-E+{A\over x^4}+{\gamma(\gamma+1)\over x^2}=0$$
implies
$$x_0=\sqrt{p+\sqrt{p^{2}+{A\over E-1}}}\eqno(4.5)
$$
where $p={\gamma (\gamma +1)\over 2(E-1)}.$ The results of our iteration method for the cases of $\alpha=1.9$ and $\alpha=2.1$ is reported in Table 1 where we compute the eigenvalues $E_{nl}^{P}$ by means of 12 iterations only. It is should be clear that these results could be further improved by increasing the number of iterations. For the case of $\alpha = 4$ we report our results in Table 2 wherein we used $x_0$ to start the iteration procedure, as given by (4.5).
\bigskip
\np %%%%%%%%%%%%% NEW PAGE KLUDGE %%%%%%%%%%%%%%%%%%%%%%
\noindent {\bf Table (I)}~~~A comparison between the exact eigenvalues $E_{nl}$ for dimension $N=2-10$ of the Hamiltonian (4.2) computed by direct numerical integration of Schr\"odinger's equation and the eigenvalues $E_{nl}^{P}$ computed by means of the present work.
\bigskip
\noindent\hfil\vbox{%
\offinterlineskip
\tabskip=0pt
\halign{\tabskip=5pt
\vrule#\strut& #\strut\hfil&\vrule#\strut&\hfil#\strut\hfil&\vrule#\strut&
\hfil#\strut\hfil&\vrule#\strut&\hfil#\strut\hfil&\vrule#\strut&\hfil#\strut\hfil&
\vrule#\strut
\tabskip=0pt\cr
\multispan2&\multispan9\hrulefill\cr
\multispan2&\multispan4\vrule\hfil$\alpha=1.9$\hfil&&\multispan3\hfil $\alpha=2.1$\hfil&\cr
\multispan2&\multispan5&\omit&\omit
\vrule\cr\noalign{\hrule}
\multispan2\vrule ${N}$&& $E_{00}$&&
		$E_{00}^{P}$&& $E_{21}$ && $E_{21}^{P}$ &\cr
\noalign{\hrule}
&2&&8.485~38&&8.485~45&&16.543~63&&16.543~76&\cr
\noalign{\hrule}
&3&&8.564~36&&8.564~42&&16.904~44&&16.904~42&\cr\noalign{\hrule}
&4&&8.795~44&&8.795~47&&17.381~71&&17.381~45&\cr
\noalign{\hrule}
&5&&9.163~09&&9.163~09&&17.955~44&&17.955~22&\cr
\noalign{\hrule}
&6&&9.646~70&&9.646~68&&18.607~07&&18.607~00&\cr
\noalign{\hrule}
&7&&10.225~04&&10.225~03&&19.320~69&&19.320~73&\cr
\noalign{\hrule}
&8&&10.879~07&&10.879~07&&20.083~41&&20.083~46&\cr
\noalign{\hrule}
&9&&11.592~98&&11.592~98&&20.885~02&&20.885~03&\cr
\noalign{\hrule}
&10&&12.354~18&&12.354~18&&21.717~61&&21.717~59&\cr
\noalign{\hrule}
}
}
\bigskip

\noindent {\bf Table (II)} A comparison between the exact eigenvalues $E_0$ for 1-dimension of the Hamiltonian (4.2) with $\alpha=4$ computed by direct numerical integration of Schr\"odinger's equation and the eigenvalues $E_{0}^{P}$ computed by means of the present work.
\bigskip

\hskip 1 true in
\vbox{\tabskip=0pt\offinterlineskip
\def\tablerule{\noalign{\hrule}}
\def\vr{\vrule height 12pt}
\halign to300pt{\strut#\vr&#
\tabskip=1em plus2em
&\hfil#\hfil
&\vrule#
&\hfil#\hfil
&\vrule#
&\hfil#\hfil
&\vrule#
&\hfil#\hfil
&\vr#\tabskip=0pt\cr
\tablerule&&$A$&&$\gamma$&&$E_0^{P}$&&$E_0$&\cr\tablerule
&&0.001&&3&&$9.00011427833$&&$9.00011427912$&\cr\tablerule
&&~&&$4$&&$11.00006349067$&&$11.00006349074$&\cr\tablerule
&&~&&$5$&&$13.00004040373$&&$13.00004040364$&\cr\tablerule
&&0.01&&$3$&&$9.00114219619 $&&$9.00114219940$&\cr\tablerule
&&~&&$4$&&$11.00063478892$&&$11.00063478889$&\cr\tablerule
&&~&&5 && 13.00040400063 && 13.00040400060&\cr\tablerule
&&0.1&& 3 && 9.01136393266 && 9.01136402618&\cr\tablerule
&&~&& 4 && 11.00633609974 && 11.00633609923&\cr\tablerule
&&~&& 5 && 11.00403643257 && 13.00403643252&\cr\tablerule
&&1 && 3 && 9.108660360401 && 9.10865860752&\cr\tablerule
&&~&& 4 && 11.06224182608 && 11.06224171938&\cr\tablerule
&&~&& 5 && 13.04001518318 && 13.04001518306&\cr\tablerule
}}

\bigskip
\np %%%%%%%%%% NEW PAGE KLUDGE %%%%%%%%%%%%%%%%%%%%
%--------------------------------
\ni{\bf 4.2 The Quartic Anharmonic Oscillator Potentials}
%--------------------------------
\medskip
\noindent We investigate the Schr\"odinger equation $H\psi=E(A)\psi$ for the quartic anharmonic oscillators \sref{\sim-\ali}, where
$$H=-{d^2\over dx^2}+x^2+Ax^4\eqno(4.6)$$
In order to obtain the energy levels using the iteration method, we write the exact wavefunctions in the form
$$\psi(x)=e^{-{x^2\over 2}}f(x)$$
Consequently, after substituting in the Schr\"odinger equation, we obtain 
$$
{d^2f\over dx^2}=2x {df\over dx}+(1-E+Ax^4)f.\eqno(4.7)
$$
or $\lambda_0(x)=2x$ and $s_0(x)=1-E+Ax^4$. We start the iteration in this case with $x_0=0,$ the value of $x$ at which the potential takes its minimum value. We report our computational results in Table 3.

 \bigskip

\noindent {\bf Table (III)}~~~A comparison between the eigenvalues $E_n$, $n=0,1,\dots,5$, for the quartic anharmonic oscillator with $A=0.1$ computed by direct numerical integration of Schr\"odinger's equation \sref{\ali} and the eigenvalues $E^{P}$ computed by means of the present work.
\bigskip 

\hskip 1 true in
\vbox{\tabskip=0pt\offinterlineskip
\def\tablerule{\noalign{\hrule}}
\def\vr{\vrule height 12pt}
\halign to200pt{\strut#\vr&#
\tabskip=1em plus2em
&\hfil#\hfil
&\vrule#
&\hfil#\hfil
&\vrule#
&\hfil#\hfil
&\vr#\tabskip=0pt\cr
\tablerule&&$n$&&\bf $E^p$&&\bf $E$ &\cr\tablerule
&&0&&$1.065286$&&$1.065286$&\cr\tablerule
&&1&&$3.306871$&&$3.306872$&\cr\tablerule
&&2&&$5.747960$&&$5.747959$&\cr\tablerule
&&3&&$8.352642$&&$8.352678$&\cr\tablerule
&&4&&$11.09835$&&$11.09860$&\cr\tablerule
&&5&&$13.96695$&&$13.96993$&\cr\tablerule
}}

\vfil\eject
% -------------------------------------------
\ni{\bf 5. Conclusion}\medskip
% -------------------------------------------
One can find a Taylor polynomial approximation about $x_0$ for an initial value problem
by differentiating the equation itself and back substituting to obtain successive values of $y^{(k)}(x_0).$ This method is perhaps as old as the very notion of a differential equation. In this paper we 
develop a functional iteration method related to this general idea and specifically taylored for an important base-class of linear equations of the form $L(y) = y^{\prime\prime} - \lambda y^{\prime} - sy = 0.$  The iteration is assumed either to terminate by the condition $s_{n}/ \lambda_{n}=s_{n-1}/\lambda_{n-1} \equiv \alpha,$ or this condition is imposed, as an approximation. 
After looking at some well-known problems which are exactly of this sort, our principle application is to Schr\"odinger eigen equations.  These latter problems are converted to the base-type by first factoring their solutions in the form $\psi(x) = f(x)y(x),$ where $f(x)$ is the large-$x$ asymptotic form, and $y$ satisfies $L(y) = 0.$ Some aspects of this approach, such as the iteration termination condition, the construction of asymptotic forms, and the choice of $x_0,$  still await more careful mathematical analysis.  However, even in its present rudimentary state, the method offers an interesting approach to some important problems. This is especially so, as it turns out, for problems such as the spiked harmonic oscillator that are known to present some profound analytical and numerical difficulties.
\bigskip
\noindent {\bf Acknowledgment}
\medskip Partial financial support of this work under Grant Nos. GP3438 and GP249507 from the 
Natural Sciences and Engineering Research Council of Canada is gratefully 
acknowledged by two of us (respectively [RLH] and [NS]). 
\np
%-------------------------------------
\references{1}
%-------------------------------------

\end